\documentclass{ws-ijmpd}

\begin{document}

\markboth{A. Bernui, B. Mota, M. J. Rebou\c{c}as and R. Tavakol}
{Large-angle anisotropy in the WMAP data}

\catchline{}{}{}{}{}

\title{A NOTE ON THE LARGE-ANGLE ANISOTROPIES \\ IN THE WMAP CUT-SKY MAPS}


\author{A. BERNUI}
\address{Instituto Nacional de Pesquisas Espaciais --
         Divis\~{a}o de Astrof\'{\i}sica  \\
         Av. dos Astronautas 1758 \\
        12227-010 S\~ao Jos\'e dos Campos -- SP, Brazil}

\author{B. MOTA, \ M.J. REBOU\c{C}AS}
\address{Centro Brasileiro de Pesquisas F\'{\i}sicas  \\
Rua Dr.\ Xavier Sigaud 150 \\ 22290-180 Rio de Janeiro -- RJ,
Brazil }

\author{R. TAVAKOL}
\address{Astronomy Unit --  School of Mathematical Sciences \\
         Queen Mary, University of London \\
         Mile End Road, London E1 4NS, UK}

\maketitle

\begin{history}
\received{Day Month Year}
\end{history}

\begin{abstract}

Recent analyses of the WMAP data seem to indicate the possible
presence of large-angle anisotropy in the Universe. If confirmed,
these can have important consequences for our understanding of the
Universe. A number of attempts have recently been made to
establish the reality and nature of such anisotropies in the CMB
data. Among these is a directional indicator recently proposed by
the authors. A distinctive feature of this indicator is that it
can be used to generate a sky map of the large-scale anisotropies
of the CMB maps. Applying this indicator to full-sky temperature
maps we found a statistically significant preferred direction. The
full-sky maps used in these analyses are known to have residual
foreground contamination as well as complicated noise properties.
Thus, here we performed the same analysis for a map where regions  
with high foreground contamination were removed. We find that the  
main feature of the full-sky analysis, namely the presence of a   
significant axis of asymmetry, is robust with respect to this 
masking procedure. Other subtler anomalies of the full-sky are on 
the other hand no longer present.
\end{abstract}


\section{Introduction}  \label{Intro}
The wealth of high resolution data provided by the Wilkinson
Microwave Anisotropy Probe (WMAP)\cite{WMAP1}\cdash\cite{WMAP2}
has confirmed to very good approximation the standard cosmological
picture, which predicts a statistically isotropic Gaussian random
cosmic microwave background (CMB) temperature fluctuations.
Despite this success, several large-scale anomalies in the CMB
have been reported including indications of non-Gaussianity,%
\cite{Copi-etal04}\cdash\cite{Non-Gauss} evidences for a
North-South asymmetry,\cite{Eriksen-etal2004a} and the so-called
``low-$\ell$ anomalies'' such as the surprisingly small values of
the CMB  quadrupole and octopole moments,\cite{WMAP-Spergel} and
the alignment of the quadrupole and octupole
moments,\cite{TOH2003}\cdash\cite{Align} (in this connection 
see Ref.~\refcite{Wiaux}) 
whose direction has been suggested to extend to the higher
multipoles\cite{Land-Magueijo2005} (see also 
Ref.~\refcite{Copi-etal05} for a detailed discussion). In
addition, there are also indications for a preferred axis of
symmetry or directions of maximum
asymmetry.\cite{Bunn-Scott00}\cdash\cite{Donoghue}

The possible origins of such unexpected anomalous features of CMB are at 
present the object of intense investigation, with several potential 
explanations, including unsubtracted foreground contamination and/or 
systematics,\cite{Schwarz-etal04} unconsidered local effects, 
\cite{Vale05+Cooray-Seto05} other mechanisms to break 
statistical isotropy,\cite{Gordon+Tomita} and also extra-galactic origin (see 
Refs.~\refcite{Eriksen-etal2004a},~\refcite{TOH2003},~\refcite{Land-Magueijo2005} 
and~\refcite{Bunn-Scott00}--\refcite{Land-Magueijo04} for details, 
and Ref.~\refcite{actual} for recent related references). 
If they turn out to have a cosmological nature, however, they could
have far reaching consequences for our understanding of the
Universe, in particular for the above-mentioned standard
cosmological scenario.

Recently we proposed\cite{BMRT2005} a new directional indicator
$\sigma = \sigma (\theta,\phi)$, based on pair angular separation
histogram (PASH),\cite{BernuiVillela2005} to measure large-angle
anisotropy in the WMAP data. An important feature of our indicator
is that it can be used to generate a sky map of large-angles
anisotropies from CMB temperature fluctuations maps. We have
produced and studied in details $\sigma-$maps generated from the
full-sky LILC,\cite{Eriksen-etal2004b} `cleaned' TOH,\cite{TOH2003} 
and co-added\cite{WMAP2} WMAP maps, and found a statistically  
significant preferred direction in these WMAP maps,
which agrees with the preferred asymmetry axes recently
reported.\cite{Eriksen-etal2004a,Land-Magueijo04} These results
were found to be robust with respect to the choice of the full-sky
WMAP CMB maps employed. 
However, since full-sky maps are known to have residual foreground 
contamination\cite{Eriksen-etal2004b} and complicated noise 
properties,\cite{WMAP2} their choice in the ``low-$\ell$'' studies is 
not a consensus.\cite{Copi-etal05,BEBG2005} Thus, the question arises 
as to whether our results hold for cut-sky maps. Our main aim here,
which extends and complements our previous work,\cite{BMRT2005} is
to address this question by considering the LILC map with a Kp2 sky
cut. To this end, in the next section we give an account of our
large-angle anisotropy indicator,  while in the last section we
apply our indicator to the LILC map with a Kp2 sky cut, and present
our main results and conclusions.

\section{Large-angle Anisotropy Indicator}  \label{Indicator}

For a detailed discussion of the indicator briefly presented 
in this section we refer the readers to Ref.~\refcite{BMRT2005}.

The key point in the construction of our indicator is that a
homogeneous distribution of points on a two-sphere $S^2$ is seen
by an observer at the center of $S^2$ as isotropically
distributed, and therefore deviations from homogeneity in this
distribution give rise to anisotropies for this observer.

Mutatis mutandis, since in CMB studies the celestial sphere is
discretized into a set of equal size pixels, with a temperature
fluctuation associated to each pixel, the idea in the CMB context
is then to construct an indicator that measures deviation from
homogeneity in the distribution of pixels with similar
temperature. The first step towards the construction of this
indicator is subdivide a given CMB map into a number of submaps,
each consisting of equal number of pixels with similar
temperatures. The next step is to devise an indicator to measure
the deviation from a homogeneous distribution of these pixels. The
construction of our indicator, $\sigma = \sigma (\theta,\phi)$ is
based on angular separation histograms (PASH), which are obtained
by counting the number of pairs of pixels whose angular separation
$\alpha$ lies within small sub-intervals (bins) $J_i \in (0,\pi]$,
of length $\delta\alpha = \pi / N_{bins}$, where
\[
J_i = \left( \alpha_i - \frac{\delta \alpha}{2} \, , \, \alpha_i +
\frac{\delta \alpha}{2}  \right] \;, \quad  i=1,2, \dots
,N_{bins}\;, \quad
\]
with the bin centers at $\alpha_i=\,(i-\frac{1}{2})\,\delta
\alpha\,$. The PASH is then defined as the following normalized
function
\begin{equation}
\label{PASH} \Phi(\alpha_i)=\frac{2}{n(n-1)}\,\,\frac{1}{\delta
\alpha} \, \sum_{\alpha \in J_i} \eta(\alpha) \; ,
\end{equation}
where $n$ is the total number of pixels in the submap, $\,\eta
(\alpha)$ is the number of pairs of pixels with separation $\alpha
\in J_i$, and where the normalization condition
$\sum_{i=1}^{N_{bins}} \Phi(\alpha_i)\,\,\delta \alpha=1\;$ holds.

Now, for a distribution of $n$ pixels in the sky sphere $S^2$ one
can compute the expected number of pairs, $\eta_{exp}(\alpha)$
with angular separation $\alpha \in J_i$. From this quantity one
obtains the normalized expected pair angular separation histogram
(EPASH), which is clearly given by
\begin{equation}
\label{defEPASH}
\Phi_{exp}(\alpha_i)=\frac{1}{N}\,\,\frac{1}{\delta \alpha}\,
              \sum_{\alpha \in J_i} \eta_{exp}(\alpha) =
\frac{1}{\delta \alpha}\,\,P(\alpha_i)= \mathcal{P}(\alpha_i) \; ,
\end{equation}
where $N= n (n-1)/2$ is the total number of pairs of pixels,
$P(\alpha_i)=\sum_{\alpha \in J_i} \eta_{exp}(\alpha) /N $ is the
probability that a pair of objects can be separated by an angular
distance that lies in the interval $J_i$,
$\,\mathcal{P}(\alpha_i)$ is the corresponding probability
density, and where the coefficient of the summation is a
normalization factor. Equation~(\ref{defEPASH}) makes it clear
that the EPASH $\Phi_{exp}(\alpha)$ gives the distribution of
probability of finding pairs of points on the sky sphere with any
angular separation 
$\alpha_i \in ( 0,\pi]$.%
\footnote{For a homogeneous and continuous distribution of points
on $S^2$ all angular separations $\,0< \alpha \leq \pi\,$ are
allowed, and the corresponding probability distribution can be
calculated to give\cite{BernuiVillela2005} $\Phi_{exp}(\alpha) =
\mathcal{P}\,(\alpha) = \frac{1}{2}\,\sin \alpha\,.$ This is the
limit of a statistically isotropic distribution of points in $S^2$
as the number of points go to infinity. One can thus quantify
anisotropy by calculating the departure of the mean observed
probability distribution $\langle \,\Phi_{obs}(\alpha_i)\,\rangle$
from it, namely $\langle \,\Phi_{obs}(\alpha_i)\, \rangle -
\Phi_{exp}(\alpha_i)\,$.}
We denote the difference between the mean PASH (MPASH), $\langle
\, \Phi_{obs}(\alpha_i) \,\rangle $, calculated from the
observational data, and the EPASH $\Phi_{exp}(\alpha_i)$, obtained
from an statistically isotropic distribution of pixels, as
\begin{equation} \label{AnisInd1}
\Upsilon (\alpha_i) \equiv \langle \, \Phi_{obs}(\alpha_i)
\,\rangle - \Phi_{exp}(\alpha_i) \;.
\end{equation}
In practice, the expected $\Phi_{exp}(\alpha_i)$ for a
statistically isotropic map is obtained simply by scrambling a CMB
map multiple times, and averaging over the resulting histograms.

Lastly, to quantify anisotropy, we distill the histogram $\Upsilon
(\alpha_i)$ into a single number, by defining the indicator
$\sigma= \sigma (\theta,\phi)$ as the variance of
$\Upsilon(\alpha_i)$ (which has zero mean), namely
\begin{equation} \label{AnisInd2}
\sigma^2  (\theta,\phi) \equiv \frac{1}{N_{bins}}
\sum_{i=1}^{N_{bins}} \Upsilon^2 (\alpha_i) \;.
\end{equation}

Calculating $\sigma$ for the whole celestial sphere would yield a
global measure of anisotropy. To obtain a directional indicator,
we can instead calculate both MPASH and EPASH for spherical-shaped 
caps. The indicator $\sigma$ can then be viewed
as a (non-local) measure of the anisotropy in the direction of the
center of the cap. Thus, by construction, $\sigma (\theta,\phi)$
measures the deviation from isotropy in a given direction, i.e.
how the observed distribution of points deviates from a
statistically isotropic one.%
\footnote{Incidentally, for a homogeneous and continuous
distribution of points on $S^2$ the EPASH for a cap of aperture
$\theta_0 \leq \pi/2$ can also be calculated in a closed
form,\cite{Teixeira03} but in practice one approximates
$\Phi^{CAP}_{\ exp}(\alpha_i)$ by the MPASH $\langle \,
\Phi^{CAP}_{\ exp}(\alpha_i) \,\rangle$ of the statistically
isotropic distribution of pixels.}

Since $\sigma$ is a discrete scalar function defined on $S^2$, it
can be expanded in spherical harmonics, and its power spectrum
$D_\ell$ can be calculated, namely
\begin{equation}
\sigma (\theta,\phi) = \sum_{\ell=0}^\infty \sum_{m=-\ell}^{\ell}
b_{\ell m} \,Y_{\ell m} (\theta,\phi) \qquad \mathrm{and} \qquad
D_\ell = \frac{1}{2\ell+1} \sum_m |b_{\ell m}|^2 \; .
\end{equation}
It then follows that if a large-angle asymmetry is present in the
CMB temperature distribution, it should significantly affect the
$\sigma-$map on the corresponding angular scales (low-$\ell$
multipoles).

In the next section, we shall generate the $\sigma-$maps from
LILC map with a Kp2 sky cut, study its main features, and make a
comparison with our previous results for the full-sky CMB
maps.\cite{BMRT2005}

\section{Main Results and Conclusions}

Given that the large-scale angular correlations are nonlocal,
$\sigma(\theta,\phi)$, calculated over a $30^{\circ}$-radius cap
centered at $(\theta,\phi)$, can be though of as a measure of the
anisotropy in the direction $(\theta,\phi)$. In our  previous
work\cite{BMRT2005} the strategy was to obtain $\sigma$ for a set
of 12,288 caps of radius $30^{\circ}$ co-centered with the same
number of pixels generated by HEALPix with $N_{side}=32$, evenly
covering the entire celestial sphere. The resulting directional
map of anisotropy was the so-called $\sigma-$map. 
We applied this new anisotropy indicator to three CMB WMAP maps:
the LILC\cite{Eriksen-etal2004b} and the TOH\cite{TOH2003} maps 
(which are two differently foreground cleaned full-sky maps resulting 
from the combination of the five frequency bands: K, Ka, Q, V, and W 
CMB maps measured by the WMAP satellite), 
and the co-added map, which is a weighted combination of the 
Q, V, and W WMAP maps. 

The resulting $\sigma-$map were found to be anisotropic. Briefly,
there is a prominent spot with very high $\sigma$ on the
southeastern corner, with a well defined maximum at $(b \simeq
115^\circ, l \simeq 235^\circ)$, which is close (by $16^{\circ}$)
to the direction recently indicated in
Ref.~\refcite{Land-Magueijo2005}. It was further shown (by a standard
spherical harmonics expansion) that the LILC $\sigma-$map deviates
from isotropy in a statistically significant way, with anomalously
high ($>95\%$ CL) dipole, quadrupole and octupole components (see
Figs. 3 and 4 of Ref.~\refcite{BMRT2005}). 
The higher components on the other hand fall
within the expected values. This clearly indicates that the LILC map
is not statistically isotropic. Finally, we noted that the
quadrupole component has a very peculiar shape, being very
symmetric around an axis slightly off the galactic North-South.
Indeed, $82\%$ of the total power in $D_2$ comes from an
axisymmetric component in the direction $(b=10^\circ, l=289^\circ)$, 
somewhat close to the axes of symmetry of the
temperature quadrupole and octupole found in Ref.~\refcite{TOH2003}
(about $24^{\circ}$ from both).

As previously mentioned, however, there is no consensus as to
whether the full sky cleaned maps available are indeed free of
significant galactic contamination. The question then arises as to
whether one  should study the full sky maps or confine the
analysis to regions where such contamination is small.

In view of the lack of consensus on how to perform the data
analysis,  here we examine the robustness of our previous results
by investigating the LILC map after the application of the Kp2
mask (hereafter the LILC-Kp2 map), which discards the temperature
fluctuations of 15.3\% of the total number of pixels, mainly
concentrated around the galactic plane.

Note that some of the caps with centers close to but outside the
Kp2 mask would still overlap with the mask itself. If the
intersection region is too large the $\sigma$ value would be
largely an artifact os the masking procedure. On the other hand,
if we were to exclude all caps were any overlapping occurs, we
would lose information on over half of the sky. To achieve a
balance, we shall disregard caps that obey either of the following
criteria:
\begin{itemize}
\item have the cap center within the Kp2 mask, and

\item have over $15\%$ of pixels within the Kp2 mask.
\end{itemize}
With this critical value of $15\%$ for the maximum number of
overlapping pixels typically the value of $\sigma$ calculated for
the same cap in both the LILC and LILC-Kp2 maps differ by less
than 10\%.

\begin{figure}[!tbh] 
\begin{center}
\includegraphics[width=6cm, height=3.3cm]{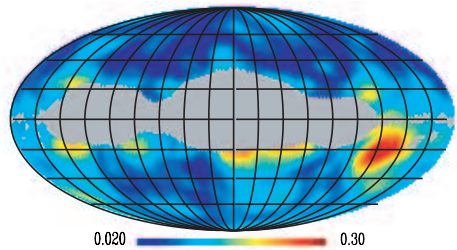} 
\end{center}
\caption{\label{FIG1} The $\sigma-$map for the LILC-Kp2 map. This
result was obtained by calculating $\sigma$ for spherical caps
with aperture $\theta_0=30^{\circ}$, following the criteria outlined 
above.} 
\end{figure}

\begin{figure}[!tbh] 
\begin{center}
\includegraphics[width=11.5cm, height=7cm]{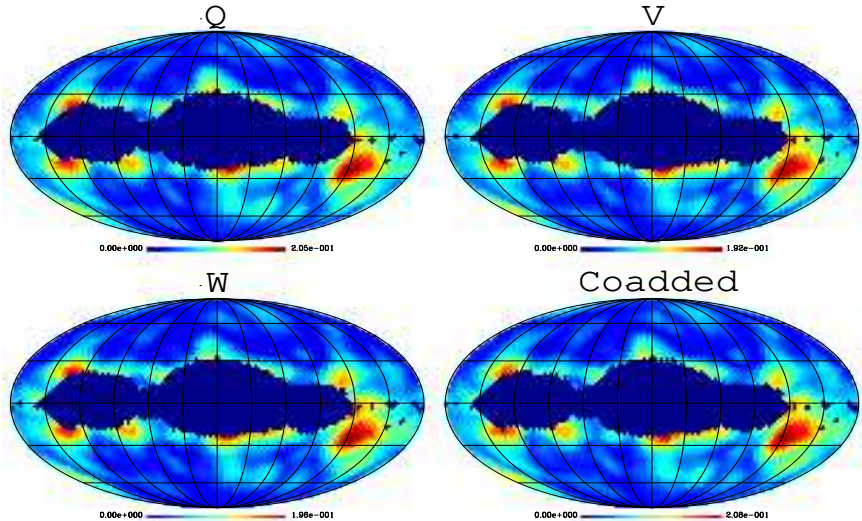}
\end{center}
\caption{\label{FIG2} The $\sigma-$map for the WMAP's Q-, V- and W-
bands, along with the combined co-added map, using the Kp2 mask.
In all maps the high $\sigma$ value is apparent.}
\end{figure}

As can be seen in Fig.~\ref{FIG1}, there is a spot of very high
$\sigma$ in the southeastern corner of the map, which coincides in
both direction and magnitude with the one found in the full 
LILC $\sigma-$map.\cite{BMRT2005} 
The fact that this spot lies well outside the region of
significant galactic contamination suggests it is not the result
of galactic contamination. The possibility remains, however, that 
it is caused by some unaccounted foreground contamination. 
Such contamination however would unlikely to affect the 
different frequency bands in exactly the same way. To verify 
whether this is the case, we calculated the $\sigma-$map for the 
Q, V and W bands separately, along with the co-added 
map\cite{Hinshaw}, which is considered the most reliable map for 
CMB studies\cite{Eriksen-etal2004b,BEBG2005} (see Fig.~\ref{FIG2}). 
The resulting maps are almost identical, supporting the result 
that foreground contamination may not account for the previously 
reported\cite{BMRT2005} large scale anisotropy.

Several new smaller high $\sigma$ spots are also in evidence near
the mask region, but they are probably just artifacts of the
masking procedure. As discussed in\cite{BMRT2005} the main
features of the resulting $\sigma-$map are robust with respect to
the number of spherical caps used to cover the celestial sphere or
the cap aperture $\theta_0$. 
                                  
\begin{figure}[!tbh]
\begin{center}
\includegraphics[width=10cm, height=6cm]{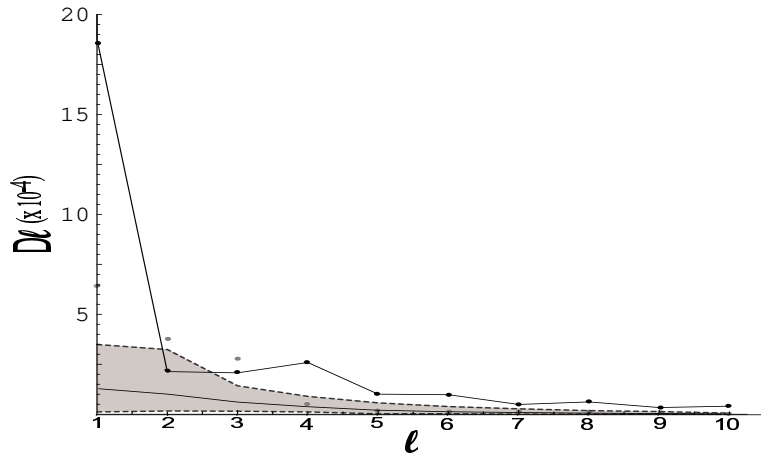}
\end{center}
\caption{\label{FIG3} The power spectrum of the LILC-Kp2 (black
dots) and LILC (grey dots) $\sigma-$maps for $\ell=1, \ldots , 10$. 
For comparison, the average $\sigma-$map power spectrum 
for a set of statistically isotropic full skies is shown (solid
curve within the grey band), along with its $95\%$ confidence 
limits (dashed curves).}
\end{figure}

To obtain more quantitative information about the observed
anisotropy, we followed the procedure of our previous work and
calculated the power spectrum of the LILC-Kp2 $\sigma-$map using
the Anafast subroutine in Healpix\cite{Gorski}. Since we are now
dealing with an incomplete sphere, the spherical harmonics are no
longer orthogonal, and the values obtained must be handled with
care. The $D_l$ values are depicted in Fig.~\ref{FIG3}, along with
the corresponding full sky values for comparison. It is clear that
the dipole component of the $\sigma-$map is even larger for the
LILC-Kp2 than for the full sky LILC. This is consistent with the
presence of an axis of asymmetry, again confirming our earlier
results. The direction of the $\sigma-$map dipole changes,
however, from $(b=141^\circ, l=240^\circ)$ in the full sky map to
$(b=150^\circ, l=209^\circ)$, in the LILC-Kp2 map, a difference of
$19^\circ$. The quadrupole and octupole components on the other
hand are comparatively smaller in the LILC-Kp2 map, probably due
to the fact that many of the high $\sigma$ structures other than the
large spot in the southeastern quadrant are now excluded 
by the Kp2 mask. 

Another feature missing from the LILC-Kp2 $\sigma-$map is the
peculiar shape of the quadrupole component observed in the full
sky map. In the latter case, $82\%$ of the total power in $D_2$ 
comes from an axisymmetric component in the direction
$(b=10^\circ, l=289^\circ)$, which may be related to the observed
alignment of the temperature quadrupole and octupole. 

We have shown that a large scale anisotropy, roughly resembling an
axis of asymmetry, remains a feature of the CMB sky even if the
regions where galactic contamination is large are not taken into 
account. Its essential dipolar nature in particular is a very
robust feature, and is consistent with other axes reported in the
literature obtained using different methods. This strongly
suggests that either the observable universe is intrinsically
anisotropic, or that there are other, subtler forms of foreground
contamination that have not yet been taken into account. 
Among the proposed explanations for the global preferred direction,  
it has been suggested that it could be due to a non-trivial topology
of the spatial section of the universe~\cite{Copi-etal05,WMAP-Spergel}
(for more details on cosmic topology see  the review articles 
Refs.~\refcite{CosmTopReviews}, and, e.g., Refs.~\refcite{TopSign,TopDetec}).  
If topology is indeed the origin,  the indicator $\Upsilon$ is 
promising in distinguishing between different topologies, as has been 
demonstrated by computer simulations in Ref.~\refcite{BernuiVillela2005}.
These are very exciting possibilities, and are worthy of further
investigation.

\section*{Acknowledgments}
We thank CNPq, PCI-CBPF/CNPq, PCI-INPE/CNPq-MCT and PPARC for the grants under
which this work was carried out. We acknowledge use of the Legacy Archive
for Microwave Background Data Analysis (LAMBDA). Some of the results in
this paper have been derived using the HEALPix package.

\end{document}